\documentclass[10pt,twocolumn,letterpaper]{article}    
\usepackage[pagenumbers]{style}
\usepackage{graphicx} 
\usepackage{stfloats}
\usepackage{placeins} 
\newcommand{\ours}{PhysTalk\xspace}
\usepackage{placeins}
\usepackage[T1]{fontenc}
\usepackage[scaled]{beramono}
\usepackage{listings}
\usepackage{listings}
\usepackage{xcolor}
\lstdefinestyle{pycode}{
  language=Python,
  basicstyle=\ttfamily\small,
  numbers=left,
  numberstyle=\tiny,
  stepnumber=1,
  numbersep=8pt,
  showspaces=false,
  showstringspaces=false,
  showtabs=false,
  frame=single,
  breaklines=true,
  tabsize=4,
}

\usepackage{times}
\usepackage{epsfig}
\usepackage{graphicx}
\usepackage{amsmath}
\usepackage{amssymb}
\usepackage{tikz}
\usepackage{graphicx}
\usepackage{amsmath}
\usepackage{amssymb}
\usepackage{booktabs}
\usepackage{textcomp}
\usepackage{comment}
\usepackage{adjustbox}
\usepackage{float}
\usepackage{multirow}

\usepackage[pagebackref,breaklinks,colorlinks]{hyperref}
\title{PhysTalk: Language-driven Real-time Physics in 3D Gaussian Scenes}
\author{
    Luca Collorone$^{*,1,2}$, \hspace{1em} Mert Kiray$^{*,2,3}$, \hspace{1em} Indro Spinelli$^1$, \hspace{1em}  Fabio Galasso$^1$, \hspace{1em} Benjamin Busam$^{2,3}$\\
{\small
\textsuperscript{1}\,Sapienza University of Rome \qquad
\small \textsuperscript{2}\,Technical University of Munich \qquad
\small \textsuperscript{3}\,Munich Center for Machine Learning (MCML) \qquad
}
  \\
  {\tt\small \{name.surname\}@uniroma1.it} \qquad
  {\tt\small mert.kiray@tum.de} \qquad
  {\tt\small b.busam@tum.de}
}

\begin{document}
\makeatletter
\g@addto@macro\@maketitle{
  \vspace{-35pt}
  \begin{figure}[H]
  \setlength{\linewidth}{\textwidth}
  \setlength{\hsize}{\textwidth}
  \centering
    \includegraphics[width=\linewidth]{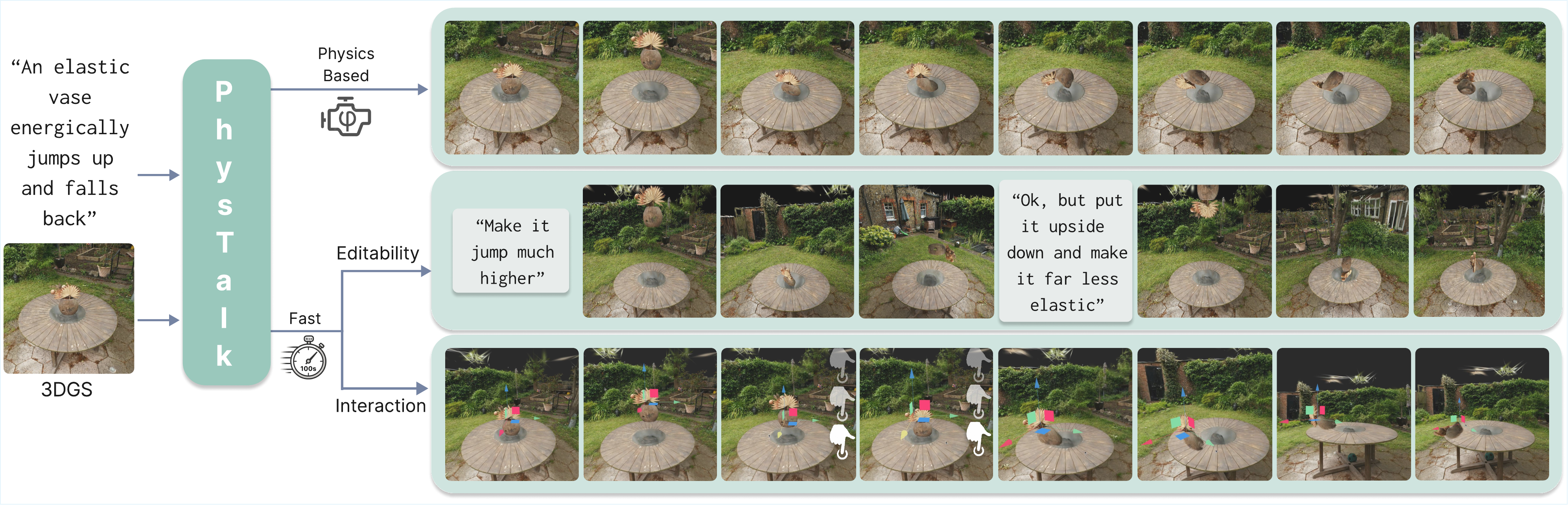}
	\vspace{-17px}
 \caption{\ours takes as input a natural language prompt together with a 3DGS scene and produces physically consistent 4D animations (first row). The efficiency of the pipeline enables rapid iteration, on the fly refinement (second row), and interactive manipulation of simulated behavior (yellow arrows in third row's samples indicate the user's push direction).}
    \label{fig:teaser}
	\end{figure}
}
\makeatother
\maketitle

\renewcommand{\thefootnote}{\fnsymbol{footnote}} 
\footnotetext[1]{Authors contributed equally.}
\setcounter{footnote}{0}
\renewcommand{\thefootnote}{\arabic{footnote}}
\begin{abstract}
\vspace{-20px}

\noindent Realistic visual simulations are omnipresent, yet their creation requires computing time, rendering, and expert animation knowledge. Open-vocabulary visual effects generation from text inputs emerges as a promising solution that can unlock immense creative potential. However, current pipelines lack both physical realism and effective language interfaces, requiring slow offline optimization.
In contrast, \textbf{\ours} takes a 3D Gaussian Splatting (3DGS) scene as input and translates arbitrary user prompts into real time, physics based, interactive 4D animations.
A large language model (LLM) generates executable code that directly modifies 3DGS parameters through lightweight proxies and particle dynamics. Notably, \ours is the first framework to couple 3DGS directly with a physics simulator without relying on time consuming mesh extraction. While remaining open vocabulary, this design enables interactive 3D Gaussian animation via collision aware, physics based manipulation of arbitrary, multi material objects.
Finally, \ours is train-free and computationally lightweight: this makes 4D animation broadly accessible and shifts these workflows from a “render and wait” paradigm toward an interactive dialogue with a modern, physics-informed pipeline.

\end{abstract}    
\section{Introduction}

Visual content creators increasingly seek tools to animate arbitrary 3D objects in a scalable, controllable, and physically realistic way. Traditional CGI pipelines require painstaking manual rigging, skeletal joints, or bespoke mesh deformation setups. They are bottlenecks in modern production. Moreover, manually authoring physically realistic motion, such as a vase falling and deforming on a table, requires expert simulation knowledge. Meanwhile, the emergence of 3D Gaussian Splatting (3DGS) has revolutionized static scene representation, but bringing the advantages of its photorealism to life with dynamic, plausible motion remains a major open challenge. The ideal system would allow a user to simply describe a desired effect such as ``\textit{make the vase jump up and fall back}'' and see it executed instantly and realistically.

Current approaches are fundamentally divided, forcing a trade-off between creative control and physical realism.
On one side, text-driven methods \cite{hsu2025autovfx, wimmer2025gaussianstolife, kiray2025promptvfx} offer impressive open-vocabulary control. They can manipulate Gaussian attributes directly from a prompt, but \emph{ignore physics}. The resulting motion is often ``floaty'' or nonsensical, lacking gravity, collisions, and material properties, failing the test of realism.
On the other side, physics-integrated Gaussians \cite{xie2024physgaussian, feng2024splashing} achieve stunning, realistic dynamics like fracturing and fluid flow. However, they lack a high-level creative interface. Scenarios must be hard-coded by experts, preventing open-vocabulary control, editability, and interactivity.
This ``great divide'' leaves a critical gap: no single system provides both intuitive control and physical guarantees.

\begin{figure*}[t]
    \centering
    \includegraphics[width=\linewidth]{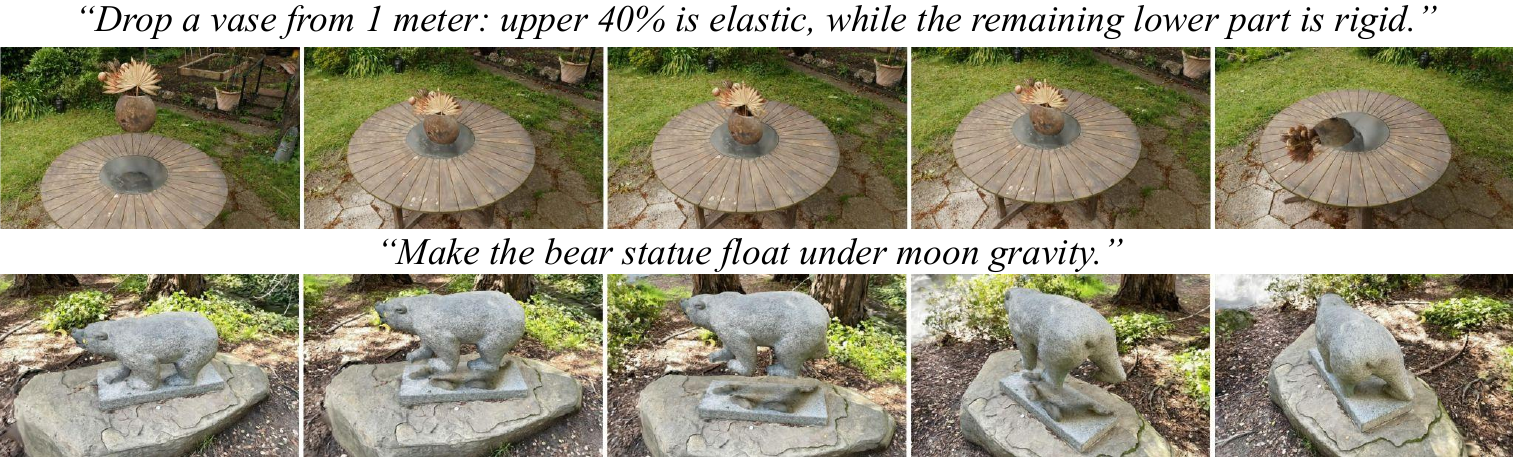}
    \vspace{-20pt}
    \caption{\ours enables a wide range of physical behaviors via \textit{Text-to-Physics Translation}. The first row shows a vase with elastic and rigid regions reacting differently when dropped: while the vase remains rigid, the impact force is transferred to the elastic flowers, which begin to wobble. The second row illustrates altered object motion under lunar gravity.
    }
    \label{fig:additional_qualitative}
\end{figure*}

We posit that this trade-off is unnecessary. Our solution is not to force a generative model to \emph{learn} the laws of physics, but to make it \emph{speak the language of physics}. We introduce \textbf{\ours}, a novel framework that bridges this gap by reframing the task as \textbf{Text-to-Physics Translation}. Our key insight is to leverage a Large-Language Model (LLM) as an intelligent ``compiler'' that translates high-level, natural language intent directly into executable, low-level code for a dedicated, high-performance physics engine such as Genesis~\cite{Genesis}.

Our pipeline is direct, fast, and object-agnostic. Given a prompt (e.g., ``\textit{make the vase jump and fall back to the table}''), our LLM generates code for the physics simulator. To connect the 3DGS representation to the engine without a complex mesh, we instantly construct a lightweight convex hull proxy from the object's Gaussian centers. The LLM-generated code then instructs the engine to run the simulation using the appropriate material model (e.g., rigid, elastic, or fluid). Finally, a skinning routine transfers the simulated motion of the physics particles back to the individual Gaussians. Because this entire process is GPU-accelerated and lightweight, the simulation runs in real-time, allowing users to edit the animation, change the camera viewpoint or interactively apply new forces to the object and see the physical response immediately (Fig.~\ref{fig:teaser}). Also, as we leverage an efficient physics engine to simulate motion, \ours has wide rage of capabilities ranging from simulations of various object materials' physical properties to simulating multi-material objects or altering gravity, as shown in Fig.~\ref{fig:additional_qualitative}. 

This ``Text-to-Physics'' approach allows \ours to uniquely satisfy all the key desiderata for a modern animation system:
\begin{itemize}
    \item \textbf{Realism:} Motion is driven by a state-of-the-art physics engine, guaranteeing adherence to gravity, collisions, and material behaviors.
    
    \item \textbf{Open-Vocabulary Simulation:} By translating free-form text into simulation code and automatically generating proxies from Gaussian centers, \ours generalizes to arbitrary prompts and objects, eliminating the need for canned effect libraries or manual rigging.
    
    \item \textbf{Interactive Physics \& Editability:} Unlike slow, mesh-based pipelines~\cite{hsu2025autovfx}, our GPU-accelerated framework runs in real-time, transforming animation into an active process. Users can iterate implicitly by re-prompting, explicitly by tweaking the generated physics code, or interactively by applying forces (e.g. pushing objects or letting them fall) with immediate visual feedback.

    \item \textbf{Multi-material Support:} Our pipeline introduces a novel hybrid coupling, enabling a single object to be simulated with heterogeneous properties from one prompt.
    
\end{itemize}

By unifying high-level language intent with low-level, interactive physics, \ours transforms 3D animation from a ``render and wait'' process into a truly ``hands-on'' creative experience.
\section{Related Work}

Our work lies at the intersection of physics-integrated 3D Gaussian Splatting (3DGS) and language-driven 3D animation. While recent efforts have explored these directions separately, their integration remains limited. We argue that \textbf{\ours} is the first framework to provide a truly \emph{expressive} and \emph{interactive} language interface for \emph{directing physics}, rather than merely assigning material properties.

\subsection{Physics-Integrated Gaussian Dynamics}

To enhance realism in 3DGS, several recent methods integrate Gaussian Splatting with physical properties \cite{jiang2024vr, qiao2023dynamic}. These approaches treat Gaussians as the medium for simulation, enabling physically grounded scene dynamics. PhysGaussian \cite{xie2024physgaussian} pioneered this direction by modeling each Gaussian as a material point within a continuum mechanics framework (MPM), achieving realistic deformations and fractures. Similarly, Gaussian Splashing \cite{feng2024splashing} coupled position-based dynamics with 3DGS to simulate rigid-body and fluid interactions. While these methods achieve high physical fidelity, their interfaces rely on graphical user interaction, where physical behaviors are manually defined.
Feature Splatting \cite{qiu2024featuresplatting} advanced this line of work by introducing predefined editing scripts and embedding rich semantic features (e.g., CLIP \cite{cherti2023openclip,radford2021learning}, DINO \cite{oquab2024dinov}) directly into Gaussians. This allows users to select objects via text (e.g., ``the vase'') and assign material properties (e.g., ``\textit{make it rigid}''). However, its use of language is fundamentally \emph{passive}: it supports property labeling and segmentation (``\textit{what is this object made of?}'') but cannot express dynamic events or causal interactions.

In summary, while physics-integrated 3DGS methods achieve impressive realism, they lack a high-level creative interface. Motions and interactions must be pre-scripted, requiring expert knowledge of simulation parameters. 
In contrast, \ours introduces a new, active paradigm. Our Text-to-Physics Translation interprets natural language as executable physics code, enabling users to freely describe complex, event-driven scenarios. This shifts the field beyond material assignment toward true, language-based direction and interactivity.

\subsection{Language-Driven VFX and Animation}

In parallel, several works have explored language as a control interface for 3D scene editing \cite{chen2024dge, chen2024gaussianeditor} and animation \cite{fang2024chat, wei2024editable, haque2023instruct}. Leveraging the rapid progress of LLM-based code generation \cite{austin2021program, drori2022neural}, recent studies have applied these ideas in computer vision \cite{gupta2023visual, lv2024gpt4motion, hu2024scenecraft, gao2024graphdreamer} and robotics \cite{singh2022progprompt, yang2024octopus}, showing that language can act as a generative interface for scene understanding and control.
The most direct approach, PromptVFX \cite{kiray2025promptvfx}, bypasses physics entirely by using an LLM to generate flow fields that deform Gaussian centers according to textual prompts. While fast and open-vocabulary, the resulting motions often lack physical realism. AutoVFX \cite{hsu2025autovfx} improves realism by producing Blender \cite{blender2018} scripts that invoke Blender’s built-in physics engine. However, this comes at the cost of expensive mesh extraction and offline rendering, limiting responsiveness and interactivity.

\textbf{\ours} builds upon the text-to-code paradigm but departs from mesh-based pipelines. By targeting a GPU-accelerated simulation framework our approach achieves real-time interactivity, object-agnostic operation, and physically consistent behavior. In doing so, it unifies the semantic flexibility of language-based control with the realism of physics-integrated 3D Gaussian Splatting.
\section{Methodology}
\begin{figure*}[t]
  \centering
  \includegraphics[width=\textwidth]{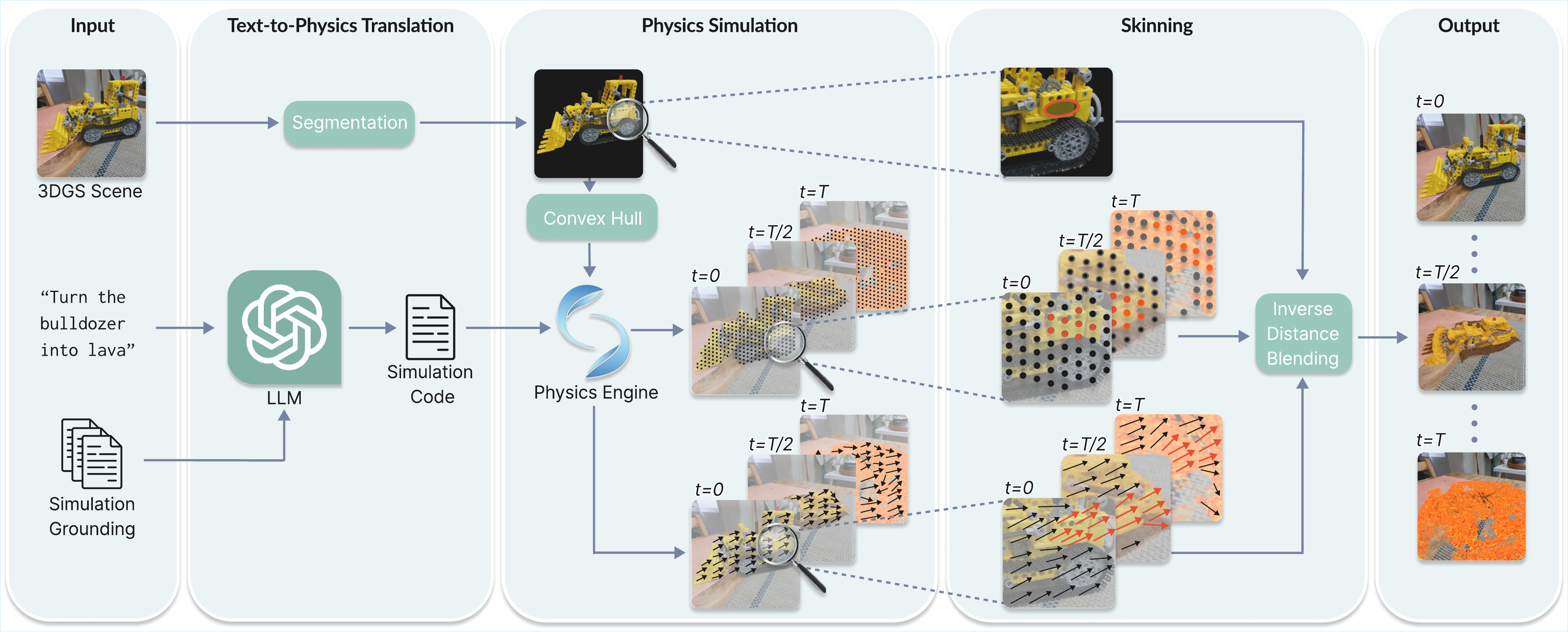}
  \caption{\textbf{\ours overview.} \textit{Input.} Our model leverages a 3DGS scene, a user prompt, and a set of simulation-grounding documents. \textit{Text-to-Physics Translation.} The text signals are fed to an LLM, which generates code for a physical simulation of the scene object. \textit{Physics Simulation.} We run the simulation and record, per frame, particle motions (black dots) and deformations (black arrows). \textit{Skinning} For each Gaussian (red circle), we select $K$ neighboring particles and their deformation gradients (red points and arrows), then apply inverse-distance blending to obtain the Gaussian’s motion and deformation; this is repeated for all Gaussians. \textit{Output.} The pipeline outputs a physics-based 4D Gaussian animation conditioned on the object and prompt.}
  \label{fig:method}
\end{figure*}
\noindent\textbf{3D Gaussian Splatting.}
3DGS models a scene as a set of anisotropic Gaussian primitives \cite{kerbl20233d}. Each Gaussian $i$ is defined by a center (position) $c_i \in \mathbb{R}^3$, a covariance matrix $\Sigma_i \in \mathbb{R}^{3\times3}$, a color $C_i$, and an opacity $\alpha_i$. The covariance $\Sigma_i$ is parameterized by a rotation quaternion $r_i$ (representing $R_i \in SO(3)$) and a scale vector $s_i$ (representing $S_i$), resulting in: 
\begin{equation}
    \Sigma_i = R_i S_i S_i^\top R_i^\top
\end{equation} A 2D-projected covariance $\Sigma_i'$ is computed via a viewing transform $W$ and projection Jacobian $J$ as $\Sigma_i' = J W \Sigma_i W^\top J^\top$ \cite{zwicker2001ewa}. The color of any pixel for \(N\) overlapping splats is then rendered via front-to-back alpha blending:
$
\mathbf{C} \;=\; \sum_{i\in N} c_i\,\alpha_i
\prod_{j=1}^{i-1}\!\big(1-\alpha_j\big),
$
where \(\mathbf{c}_i\) and \(\alpha_i\) denote the color and opacity derived from the
Gaussian parameters.

\noindent\textbf{Unified Physics Simulation.} To support open-vocabulary prompts ranging from ``push'' to ``melt,'' our framework relies on Genesis \cite{Genesis}, a multi-physics backend capable of simulating rigid bodies, elastic materials (MPM), and fluids (SPH) within a unified domain. Crucially, to drive the Gaussian skinning described in Sec. \ref{Sec:skinning}, the engine must provide access to the full deformation gradient $F_i^{(t)} \in \mathbb{R}^{3\times3}$ for each particle $i$ at every time step $t$. This tensor implicitly captures the local rotation and stretch components required to realistically deform anisotropic 3D Gaussian primitives, independent of the specific underlying solver.

\subsection{Pipeline Overview}
Our \ours pipeline transforms a static 3DGS object into a physically-plausible animation, guided entirely by a natural language prompt. \ours operates on a pre-segmented set of Gaussians corresponding to the target object, obtained either manually or through external automatic segmentation. 
As shown in Fig. \ref{fig:method}, our process is training-free and consists of three main stages:
\begin{enumerate}
    \item \textbf{Text-to-Physics Translation:} \ours leverages an LLM to convert the user's prompt into executable Python code for the Genesis physics engine.
    \item \textbf{Physics Simulation:} the generated code creates a lightweight proxy (a convex hull) from the 3DGS centers and runs the physics simulation, recording particle trajectories $p_i^{(t)}$ and deformation gradients $F_i^{(t)}$.
    \item \textbf{Gaussian Skinning:} a lightweight skinning routine transfers the simulated particle motion back to the original Gaussians, updating their centers $c_j^{(t)}$ and covariances $\Sigma_j^{(t)}$ for each frame to produce the final animation.
\end{enumerate}

\noindent Finally, opacity and color changes can be efficiently synthesized and applied by leveraging \cite{kiray2025promptvfx}. 

\subsection{Text-to-Physics Translation}
\label{sec:t2phy}
The central component of our approach is the Text-to-Physics Translation module, responsible for converting natural language prompts into executable simulator code. Because current large language models (LLMs) are not trained on specific simulators such as Genesis \cite{Genesis} and thus cannot natively produce syntactically or semantically valid code for it, we employ an in-context learning (ICL) framework to steer the model toward simulator-compliant outputs.

\noindent\textbf{1. Constrained Prompting.} We do not ask the LLM to generate code from scratch. Instead, we provide it with a structured simulation template that it must fill in. This template defines the necessary functions: a \texttt{build\_scene} function that computes the object’s convex hull, instantiates the scene and object(s) with selected material(s), and sets simulation parameters; a \texttt{step} method that advances physics, samples current particle positions and per particle deformations, and stores them in a buffer; and a \texttt{query} method that concatenates the buffered samples and performs skinning. This constrained approach dramatically improves output reliability by forcing the LLM to focus on parameter selection (e.g., material type, forces) rather than code structure.

\noindent\textbf{2. Few-Shot Exemplars.} We provide the LLM with the official physics engine API specifications and a small set of hand-written simulation functions as few-shot exemplars. These examples allow the LLM to condition the generation with the correct syntax and common patterns.

\noindent\textbf{3. Custom API and Instruction.} Additionally, we expose a small API suite for deformation estimation and particle-to-Gaussian motion skinning that the model can invoke. This is a crucial design choice: the LLM does not need to know how to perform SVD or k-d tree skinning; it only needs to call our pre-defined functions. This setup allows to steer the LLM to produce executable simulation code that is both complex and correct. Finally, we engineer a set of text instruction to guide the model to produce effects, provide executable code and avoid instable physical setups.  We report exemplars and instructions in the Supplementary Material.

\subsection{Physics Simulation and Skinning}
\label{sec:phy_and_ski}
\noindent\textbf{Proxy Generation.} Our pipeline is object-agnostic as any generated simulation's code can be applied to any object. Given a 3DGS object $\mathcal{G}$, we approximate the object by wrapping its points with a lightweight convex hull derived from its Gaussian centers $\mathbf{c}_{j}$. This proxy is coarse enough to keep setup and simulation fast, yet expressive enough to support accurate physics.

\noindent\textbf{Simulation.} The LLM-generated \texttt{build\_scene} function instantiates this convex hull in the simulator, discretizes it into $N$ simulation particles $\mathbf{P}^{(t=0)}=\{\mathbf{p}_{i}^{(t=0)}\}_{i=1}^{N}$, and assigns the chosen material properties. The \texttt{step} function then runs the simulation, recording the world-space displacement $\mathbf{d}_{i}^{(t)}=\mathbf{p}_{i}^{(t)}-\mathbf{p}_{i}^{(0)}$ and the per-particle deformation gradient $\mathbf{F}_{i}^{(t)}\in\mathbb{R}^{3\times3}$ for each particle at each frame.
This gradient $\mathbf{F}_i^{(t)}$ fully describes the local deformation, as motivated by its Singular Value Decomposition (SVD):
\begin{equation}
\mathbf{F}_i^{(t)} = \mathbf{U}_i^{(t)} \, \mathbf{S}_i^{(t)} \, \mathbf{V}_i^{(t)\top},
\end{equation}
where $\mathbf{U}_i^{(t)}, \mathbf{V}_i^{(t)} \in \mathrm{O}(3)$ and $\mathbf{S}_i^{(t)} = \mathrm{diag}(s_{i,1}^{(t)}, s_{i,2}^{(t)}, s_{i,3}^{(t)})$. This decomposition allows the rotation and stretch components to be represented respectively as:
\begin{equation}
\mathcal{R}_i^{(t)} = \mathbf{U}_i^{(t)} \mathbf{V}_i^{(t)\top}, \qquad 
\mathcal{S}_i^{(t)} = \mathbf{V}_i^{(t)} \mathbf{S}_i^{(t)} \mathbf{V}_i^{(t)\top}.
\end{equation}

\noindent\textbf{Gaussian Skinning.} \label{Sec:skinning} To transfer this motion to the Gaussians, we first associate each Gaussian $g_{j}$ with a fixed set of $K$ nearby particles $\mathcal{N}_{j}$ based on a $k$-d tree query at the rest pose. We then compute the skinning weights $w_{j,i}$ for each neighbor particle $i \in \mathcal{N}_j$ based on their inverse distance at the rest pose, ensuring $\sum_{i \in \mathcal{N}_j} w_{j,i} = 1$:
\begin{equation}
w_{j,i} = \frac{\bigl( \lVert \mathbf{c}_j - \mathbf{p}_i^{(0)} \rVert_2^2 + \epsilon \bigr)^{-1}}
{\sum_{k \in \mathcal{N}_j} \bigl( \lVert \mathbf{c}_j - \mathbf{p}_k^{(0)} \rVert_2^2 + \epsilon \bigr)^{-1}}
\end{equation}
where $\mathbf{c}_j$ is the Gaussian center, $\mathbf{p}_i^{(0)}$ is the particle rest position, and $\epsilon$ is a small constant for numerical stability.

 For every step $t$, the \texttt{query} function updates the Gaussian center $\hat{\mathbf{c}}_{j}^{(t)}$ by adding a weighted sum of its neighbors' displacements to its original center $\mathbf{c}_{j}$:
\begin{equation}
\hat{\mathbf{c}}_{j}^{(t)}=\mathbf{c}_{j}+\sum_{i\in \mathcal{N}_{j}}w_{j,i}\mathbf{d}_{i}^{(t)}
\end{equation}
Similarly, the effective deformation gradient for the Gaussian, $\hat{\mathbf{F}}_{j}^{(t)}$, is a weighted sum of the particle deformation gradients:
\begin{equation}
\hat{\mathbf{F}}_{j}^{(t)}=\sum_{i\in\mathcal{N}_{j}}w_{j,i}\mathbf{F}_{i}^{(t)}
\end{equation}
This blended deformation gradient $\hat{\mathbf{F}}_j^{(t)}$ implicitly averages the local rotations $\mathcal{R}_i$ and stretches $\mathcal{S}_i$ from the particle neighborhood. We apply this transformation directly to the original Gaussian covariance $\mathbf{\Sigma}_{j}$ using the standard transformation:
\begin{equation}
\hat{\mathbf{\Sigma}}_{j}^{(t)}=\hat{\mathbf{F}}_{j}^{(t)}\mathbf{\Sigma}_{j}(\hat{\mathbf{F}}_{j}^{(t)})^{\top}
\end{equation}
This lightweight, parallelizable process ``skins" the physical simulation onto the 3DGS object, ultimately producing a realistic, dynamic animation. 

\noindent \textbf{Generalization to Continuum Materials.} To generalize animation to fluids, we leverage smoothed-particle hydrodynamics (SPH) descriptions without per-particle deformation gradient by setting $\hat{\mathbf{F}}_{j}=\mathbf{I}$. Note that, as flow expands on a surface, isolated Gaussians can drift apart and visible gaps can emerge. We address this by detection of holes whenever inter-Gaussian spacing increases significantly. For each hole, we spawn new Gaussian centers on multiple radial shells if a Poisson-disk distance test is satisfied. This targeted insertion fills gaps and maintains visual continuity even for fluid streams. We ablate this approach in the Supplementary Material. 

\section{Experiments}
\label{sec:experiments}

\subsection{Experimental Details}

\noindent\textbf{Physics Engine.} We instantiate our framework using Genesis~\cite{Genesis}, an open-source, GPU-accelerated physics engine selected for its high performance and flexible Python API. Genesis supports rigid objects, continuum materials, and particle-based fluids in a single environment. 

\noindent \textbf{Dataset \& Preprocessing.} We evaluate our method on four real-world scenes: the \textit{garden vase} and \textit{bulldozer} scenes from Mip-NeRF360~\cite{barron2022mipnerf360}, the \textit{bear} scene from Instruct-NeRF2NeRF~\cite{instructnerf2023}, and the \textit{horse} scene extracted from Tanks and Temples~\cite{Knapitsch2017}. 
These scenes are reconstructed using Gaussian Splatting, yielding a set of 3D Gaussians that jointly encode the geometry and color of the scene and its objects. 

\noindent \textbf{Baselines.} We compare our approach against three baselines that generate 4D Gaussians from text. In particular, we consider Gaussians2Life \cite{wimmer2025gaussianstolife}, which uses a text conditioned video diffusion model to synthesize a 2D motion that is subsequently lifted to 4D Gaussians via an offline optimization stage. Also, we employ AutoVFX \cite{hsu2025autovfx}, which translates prompts into Blender scripts, converts the 3DGS object into a mesh, run the computation and then renders the result: this requires expensive mesh extraction and offline rendering. Both pipelines involve heavy preprocessing and do not support user interaction or editing. In addition, we employ PromptVFX \cite{kiray2025promptvfx}, which applies LLM generated transformation fields directly to Gaussians in real time, providing fast text control at the cost of sacrificing explicit physical grounding. 

\noindent \textbf{Implementation Notes.} Our method is implemented in PyTorch and uses GPT-5 \cite{achiam2023gpt} to translate text prompts into Genesis simulation code. All experiments are run on a single NVIDIA RTX 4090 GPU, although we observed that the simulation consistently uses under 4 GB of VRAM, enabling the method to run on consumer-grade GPUs as well. We use \(K = 8\) nearest simulation particles per Gaussian when applying the skinning procedure described in Sec.~\ref{Sec:skinning}.

\noindent\subsection{Qualitative Evaluation} 

\begin{figure*}[p]
    \centering
    \includegraphics[width=\linewidth]{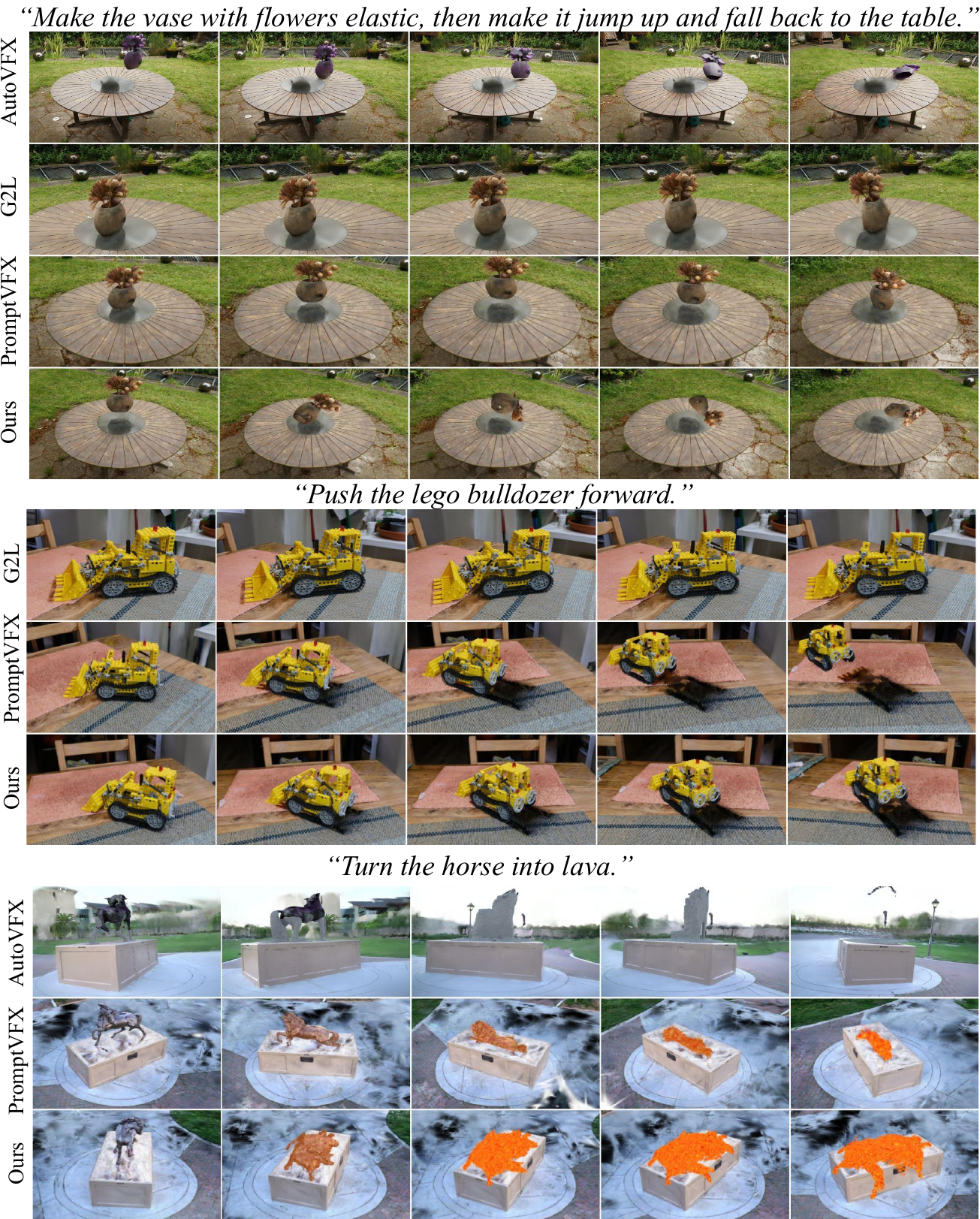}
    \vspace{-20pt}
    \caption{Qualitative comparison of \ours with existing baseline methods across multiple scenes and prompts. First example highlights \ours's ability to faithfully execute user prompts and generate elastic, deformable materials. Second example shows our pipeline applying a realistic push to the bulldozer. Third example shows \ours's compelling performance at fluid simulation. }
    \label{fig:qualitative_comparison}
\end{figure*}

In Fig.~\ref{fig:qualitative_comparison} we qualitatively compare all baselines and \ours on a set of three prompts. For each prompt, our pipeline generates animations by updating Gaussian parameters according to the simulation.

\noindent \textbf{Elastic Jump.} For this prompt, \ours correctly attributes elastic behavior to the vase and deforms it after impact. Contrarily, PromptVFX and AutoVFX both produce essentially rigid vases, with PromptVFX generating an unrealistically jump and AutoVFX producing a mostly ballistic fall with no visible deformation. Gaussians2Life manages to trigger a small jump but introduces noticeable non-physical distortions of the vase geometry.
 
\noindent \textbf{Push Forward.} Both \ours and PromptVFX produce plausible and visually coherent pushing motions, while Gaussians2Life fails to generate a meaningful interaction. AutoVFX is omitted from this comparison as for this prompt it yields a static result with no motion.

\noindent \textbf{Lava Flow.} This prompt tests fluid style dynamics. \ours generates a lava stream that remains dense and flows smoothly over the scene, capturing coherent fluid motion while dripping down from the pedestal. In contrast, AutoVFX produces a sparse, low density violet fluid that lacks the appearance of molten lava. PromptVFX yields a lava like material whose motion is not consistent with realistic fluid dynamics. Gaussians2Life relies on 2D diffusion based flow fields lifted to 4D Gaussians for object deformation and does not support particle based or fluid like motion. Thus, it is unable to produce a meaningful result for this prompt.

\begin{table*}[t!]
  \centering
  \caption{
    Comparison of methods across CLIP similarity, VQAScore, User Study ratings, and Runtime. 
    User Study includes Text Alignment and Animation Quality scores (1–5 Likert scale, normalized). 
    Bold indicates best per prompt and metric. Values in parentheses denote the animation duration. * For fluids we report both the runtime without / with our hole-filling strategy. 
  }
  \label{tab:method-comparison}
  \vspace{-10pt}
  \resizebox{2\columnwidth}{!}{%
    \begin{tabular}{l l c c ccc}
      \toprule
      \textbf{Setup} & \textbf{Method} 
      & \textbf{CLIP~\cite{radford2021learning}} 
      & \textbf{VQAScore~\cite{lin2024evaluating}} 
      & \multicolumn{2}{c}{\textbf{User Study}} 
      & \textbf{Runtime (s)} \\
      \cmidrule(lr){5-6}
      & 
      & 
      & 
      & \textbf{Text Alignment} 
      & \textbf{Animation Quality} 
      & \\
      \midrule
      \multirow{4}{*}{Elastic vase jumps and falls. (2s)}
        & AutoVFX~\cite{hsu2025autovfx}                  & \textbf{0.2223} & 0.6069 & 0.2295$\pm$0.24 & 0.3928$\pm$0.29 & 4149 \\
        & Gaussians2Life~\cite{wimmer2025gaussianstolife} & 0.2042 & 0.3121 & 0.1683$\pm$0.23 & 0.2193$\pm$0.28 & 490 \\
        & PromptVFX \cite{kiray2025promptvfx}                                         & 0.1888 & 0.2417 & 0.3316$\pm$0.25 & 0.3316$\pm$0.26 & 100 \\
        & \ours                                           & 0.1908 & \textbf{0.6791} & \textbf{0.7704$\pm$0.22} & \textbf{0.4234$\pm$0.28} & \textbf{96} \\
      \midrule
      \multirow{4}{*}{Push the bulldozer forward. (1s)}
        & Gaussians2Life~\cite{wimmer2025gaussianstolife} & \textbf{0.2821} & 0.7141 & 0.1785$\pm$0.21 & 0.2040$\pm$0.23 & 331 \\
        & PromptVFX \cite{kiray2025promptvfx}                                        & 0.2014 & 0.7589 & \textbf{0.8367$\pm$0.23} & 0.5408$\pm$0.28 & \textbf{70} \\
        & \ours                                           & 0.2035 & \textbf{0.7769} & 0.8265$\pm$0.19 & \textbf{0.6326$\pm$0.30} & 90 \\
      \midrule
      \multirow{4}{*}{Turn the horse into lava. (2s)}
        & AutoVFX~\cite{hsu2025autovfx}                  & 0.1588 & 0.2548 & 0.3418$\pm$0.25 & 0.3316$\pm$0.25 & 5874 \\
        & PromptVFX \cite{kiray2025promptvfx}                                      & \textbf{0.2030} & 0.3203 & 0.7755$\pm$0.24 & 0.4693$\pm$0.27 & \textbf{61}\\
        & \ours                                           & 0.1997 & \textbf{0.5244} & \textbf{0.8673$\pm$0.17} & \textbf{0.5918$\pm$0.30} & 63 / 1238$^{*}$ \\
      \bottomrule
    \end{tabular}
      }
\end{table*}

\subsection{Quantitative Evaluation}

\noindent \textbf{Metrics.} Benchmarking 4D Gaussian animation is challenging, since no ground truth sequences are available. Following recent work~\cite{chen2024dge,hsu2025autovfx, wimmer2025gaussianstolife, kiray2025promptvfx}, we report CLIP similarity~\cite{radford2021learning} as a proxy for text to single-image alignment. However, CLIP is computed at the frame level, so it does not capture motion quality, physical realism, or text animation coherence, all aspects that can only be appreciated when viewing the animation over time \cite{kiray2025promptvfx}.
To better account for video level consistency, we follow \cite{kiray2025promptvfx} and additionally report the VQAScore~\cite{lin2024evaluating}, a video based metric that has been shown to correlate more strongly with human judgment. The VQAScore estimates the probability that a vision language model answers “Yes” to the question \emph{Does this video align with the described animation: `\{prompt\}''?}, denoted as $\mathbb{P}(\text{`Yes''} \mid \text{video, prompt})$.

As a complement to these metrics we also conduct a user study to assess the perceived quality of the generated animations. We collect preferences from 50 human evaluators and follow the protocol introduced by \cite{hsu2025autovfx}, asking participants to rate each sample along two dimensions, \textit{Text Alignment} and \textit{Animation Quality}. Ratings are given on a 1–5 Likert scale and subsequently normalized to the range [0, 1] using min-max normalization.

Finally, since our goal is to design a low-latency pipeline that can be effectively used for iterative editing and interaction, we report runtimes for all compared methods. 
For fluid simulations we additionally report \ours runtime both with and without the hole filling strategy described in Sec.~\ref{Sec:skinning}. 

\begin{figure}[t]
    \centering
        \begin{subfigure}[b]{0.49\linewidth}
        \centering
        \includegraphics[width=\linewidth]{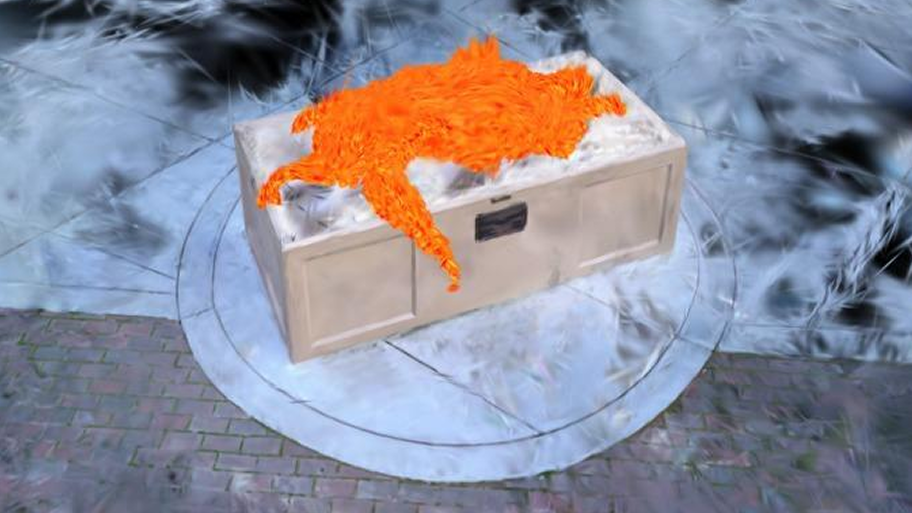}
        \caption{}
        \label{fig:lava_denser}
    \end{subfigure}
    \hfill
    \begin{subfigure}[b]{0.49\linewidth}
        \centering
        \includegraphics[width=\linewidth]{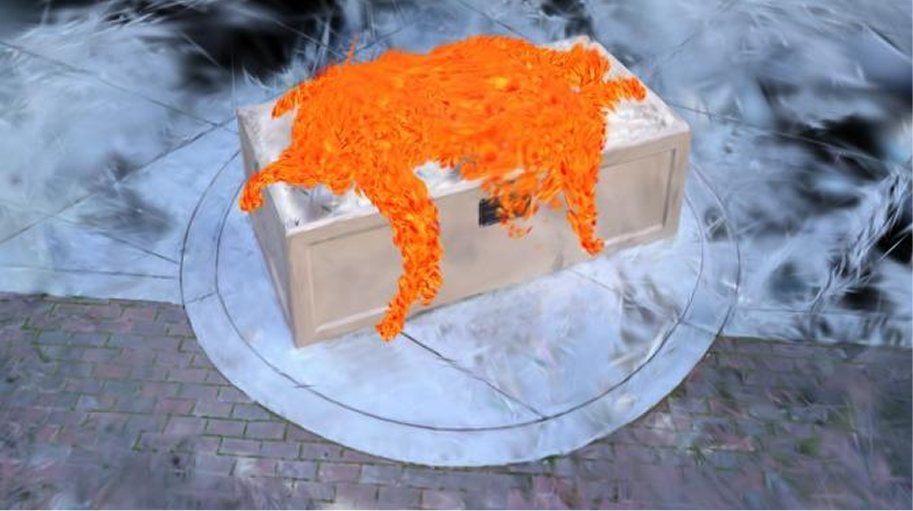}
        \caption{}
        \label{fig:lava_less_dense}
    \end{subfigure}
    \vspace{-8px}
    \caption{Comparison of lava simulations with different surface-tension parameter values.}
    \label{fig:lava_density_comparison}
\end{figure}

\noindent \textbf{Results.} Table~\ref{tab:method-comparison} shows that \ours consistently outperforms all other methods across all prompts in terms of VQAScore, while CLIP similarity remains inconclusive. Moreover, the user study indicates that human evaluators systematically prefer \ours, with perceived cumulative quality improvements of 28\% in \textit{Text Alignment} and 22\% in \textit{Animation Quality} over the best competitor, PromptVFX.

Note that \ours achieves runtimes that are comparable with the fastest competitor, PromptVFX, despite producing physically realistic, simulation-based outputs. In contrast, Gaussian2Life and, in particular, AutoVFX are much slower, with runtimes on the order of minutes rather than seconds.

\begin{figure*}[t]
\centering
\setlength{\tabcolsep}{2pt}
\renewcommand{\arraystretch}{1.1}
\begin{tabular}{@{}ccc@{}}
\includegraphics[width=0.33\textwidth]{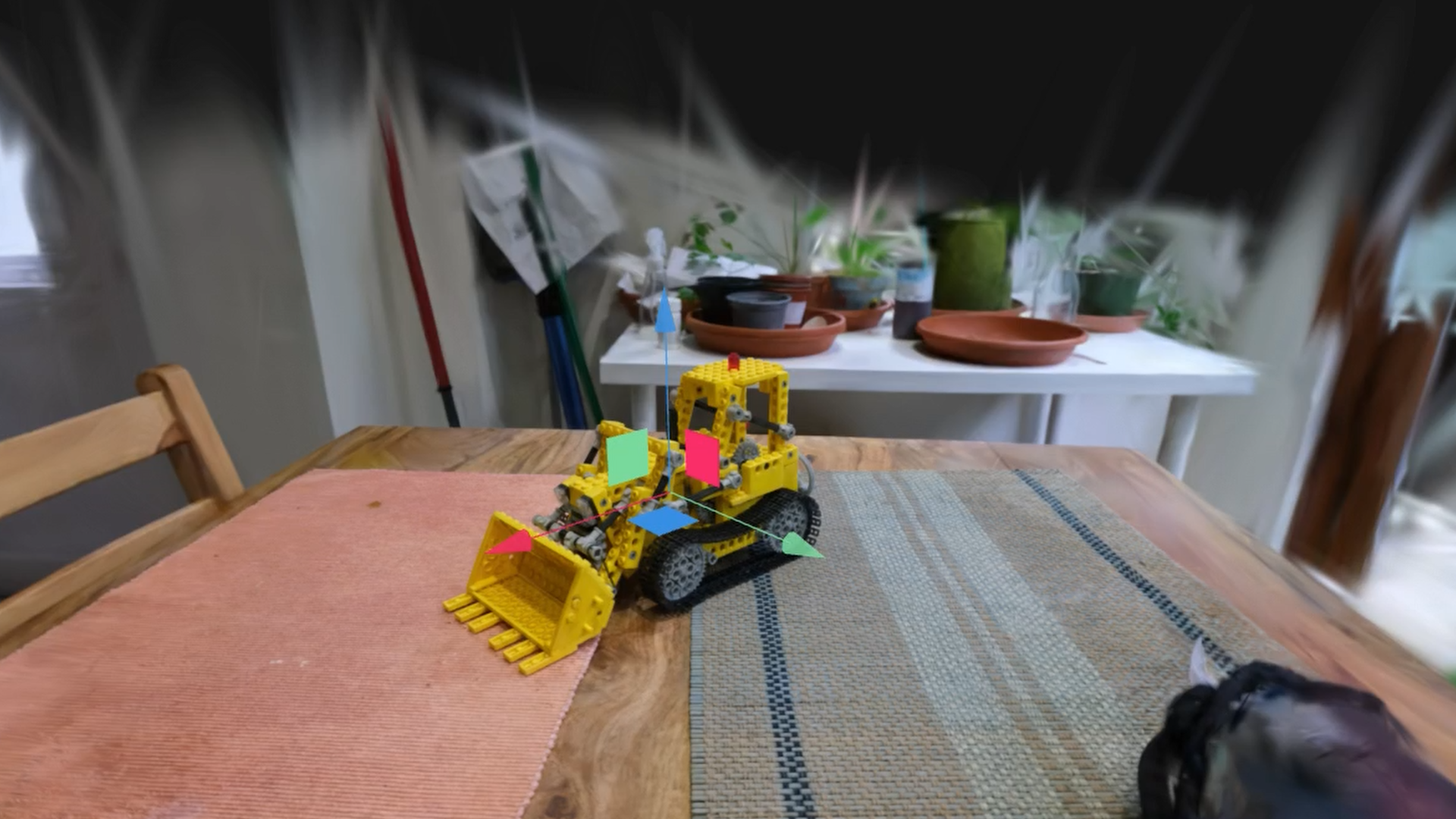} &
\includegraphics[width=0.33\textwidth]{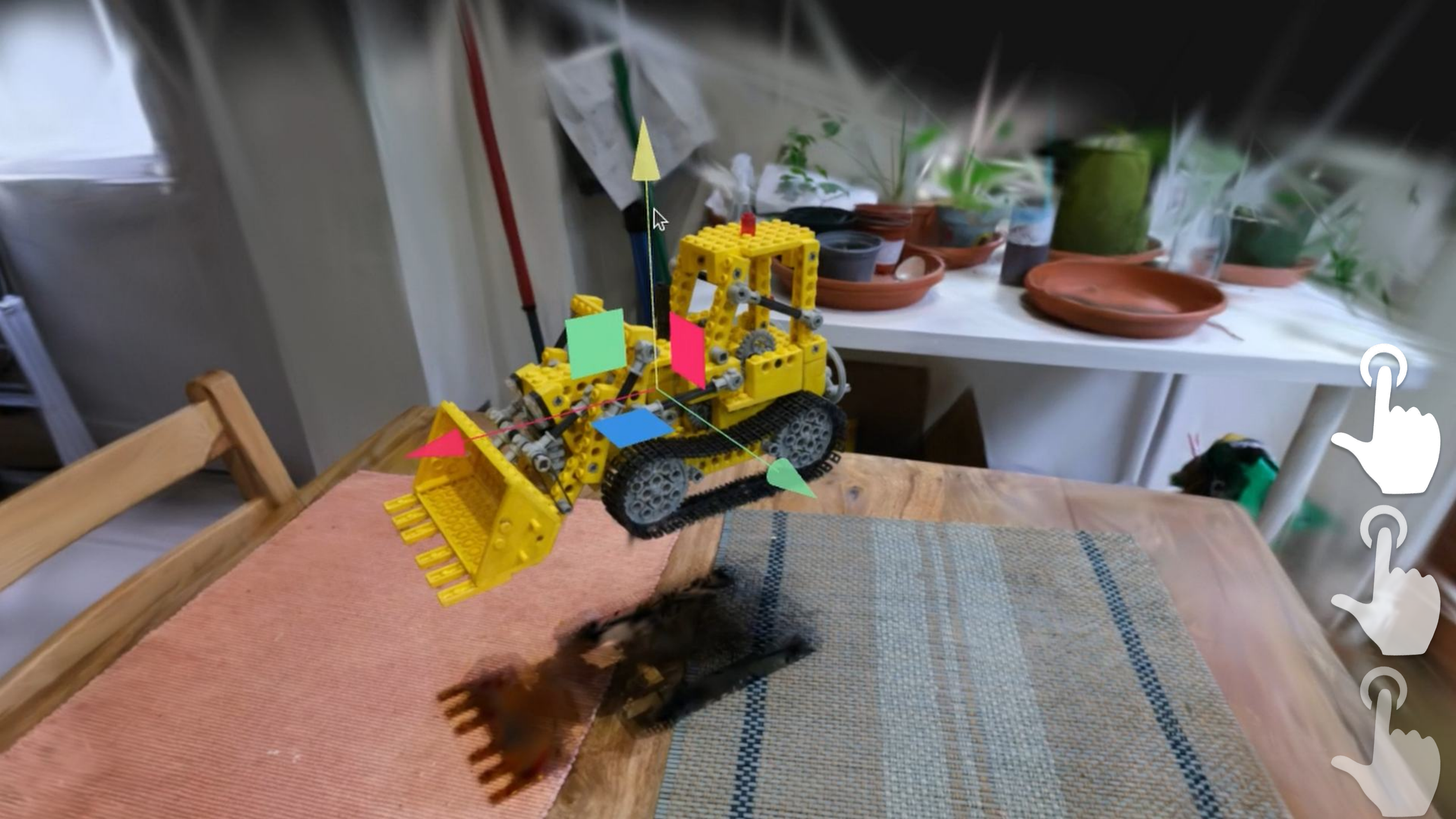} &
\includegraphics[width=0.33\textwidth]{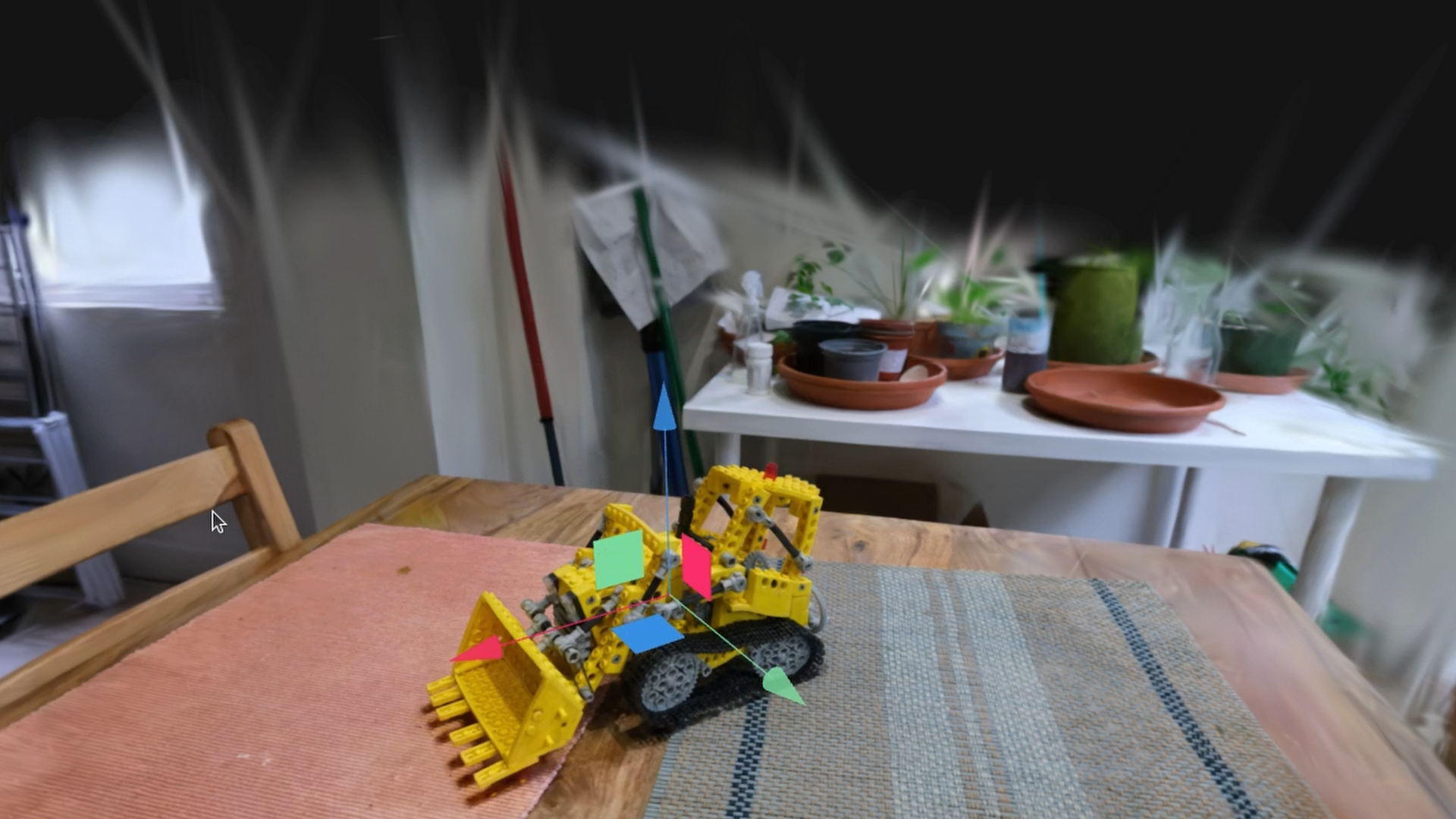} \\
\scriptsize (1) Initial State: elastic bulldozer &
\scriptsize (2) User applies an upward push &
\scriptsize (3) Reaction: the object falls \\[0.5ex]
\includegraphics[width=0.33\textwidth]{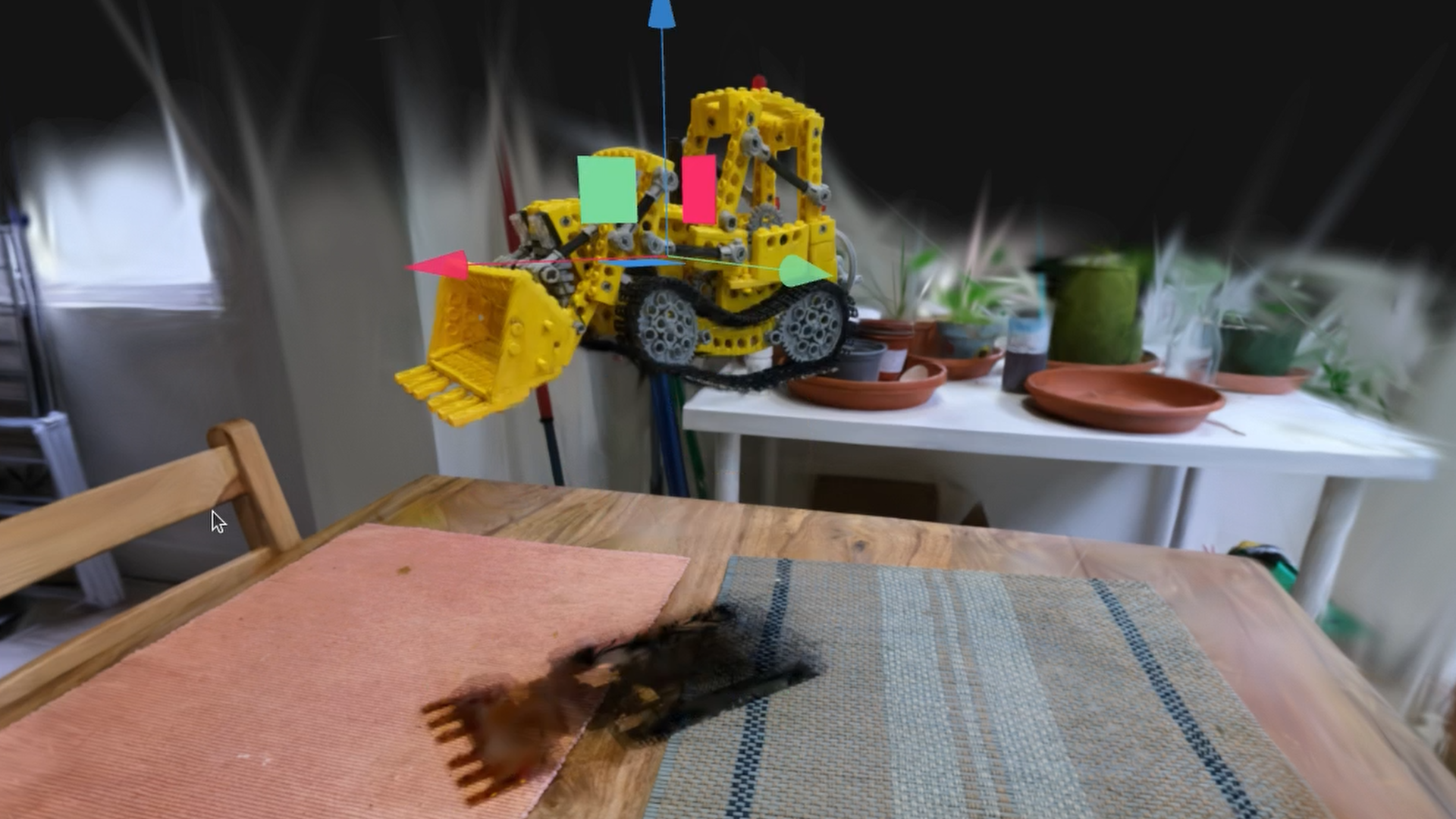} &
\includegraphics[width=0.33\textwidth]{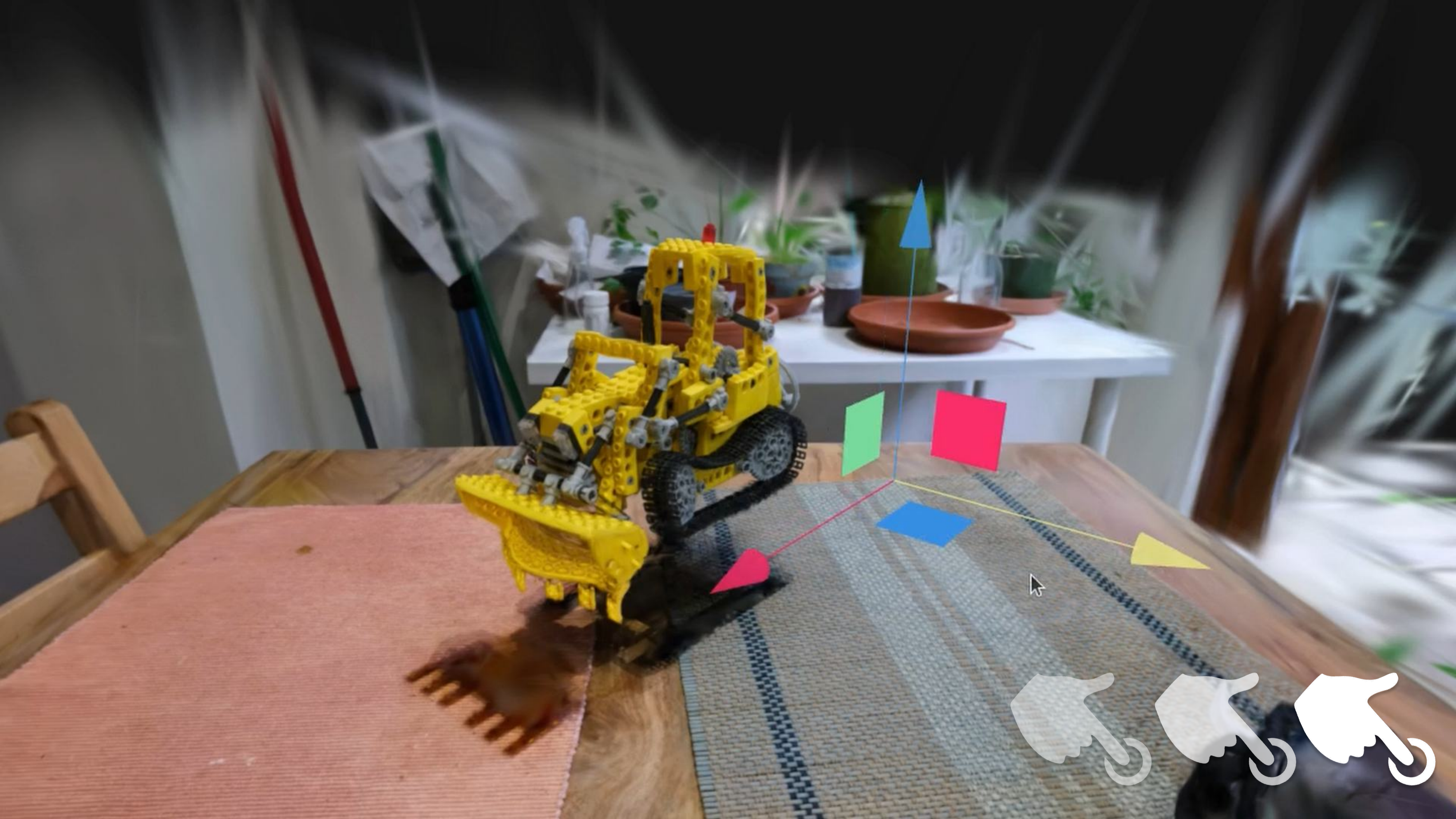} &
\includegraphics[width=0.33\textwidth]{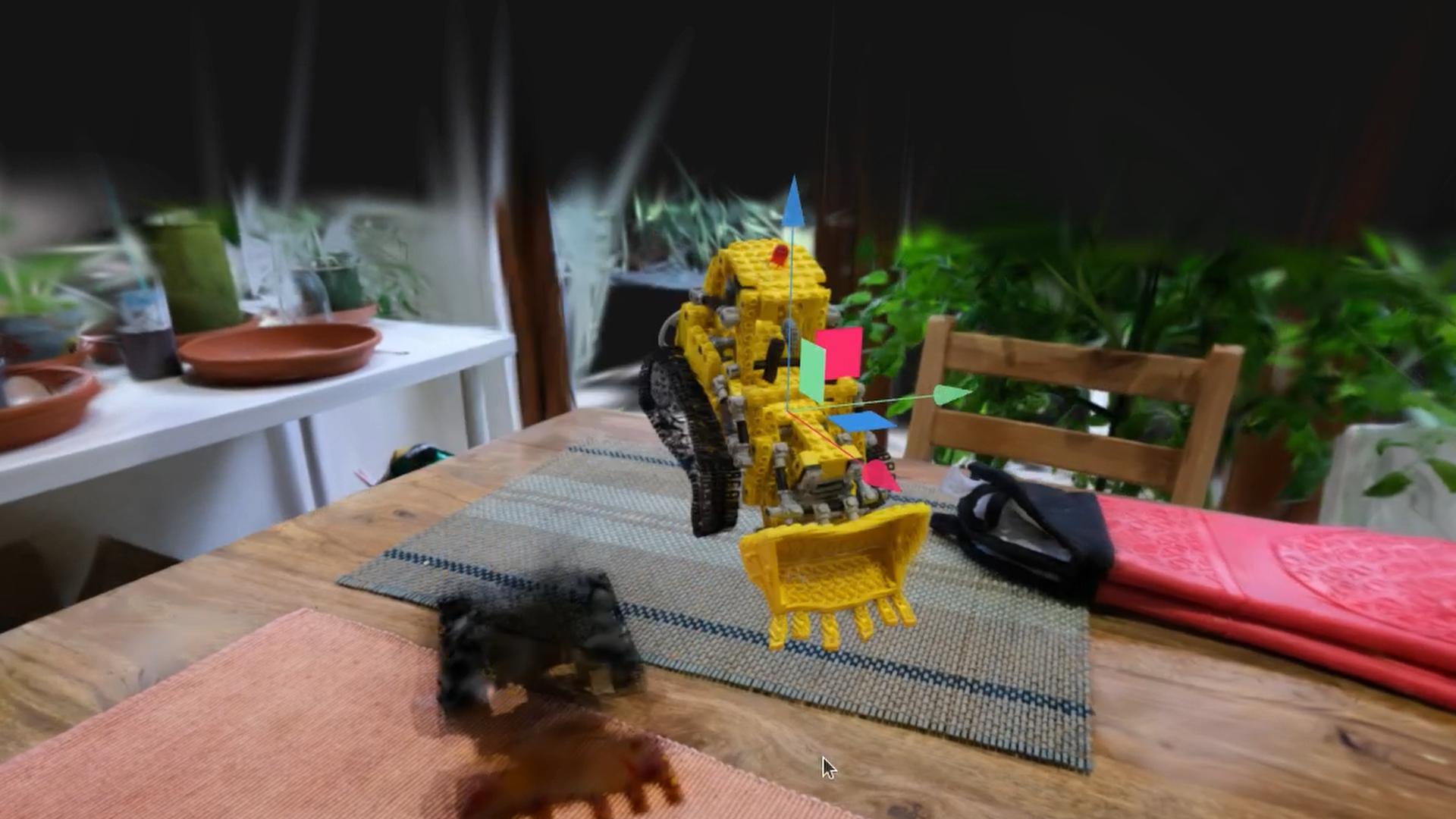} \\
\scriptsize (4) Reaction: the object bounces up &
\scriptsize (5) User applies a lateral push &
\scriptsize (6) Reaction: the object moves to the right\\
\end{tabular}
\caption{User interaction with an elastic bulldozer. Yellow arrows indicate active user-applied forces, while red, blue, and green arrows denote inactive manipulation handles. The interface allows the object to be pushed or dragged during the simulation while remaining governed by the underlying physics.}
\label{fig:bulldozer_interaction}
\end{figure*}

\subsection{Editability  \&  Interaction}

We design our framework to be fast and efficient, enabling users to refine a simulation until they are satisfied with the result. These edits can occur either implicitly, by asking the LLM for refinements, or explicitly, by tuning simulation variables. We showcase implicit edits in Fig.~\ref{fig:teaser}, where the user starts from an elastic-vase jumping simulation and iteratively modifies its behavior. Also, we illustrate explicit simulation's parameter editing: in Fig.~\ref{fig:lava_denser} the user increases the surface-tension parameter to obtain a visually denser flow compared to the original sample in Fig.~\ref{fig:lava_less_dense}.

In addition, our pipeline allows users to create simulations with specific physical behavior and interact with them at runtime in the browser. Fig.~\ref{fig:bulldozer_interaction} shows a user dragging and pushing an object, first upward and then sideways, while it remains subject to the residual forces and deformations applied throughout the animation. The interaction speed depends on the number of Gaussians and the chosen material, with observed interactive frame rates between 4 and 9 FPS.
\section{Discussion \& Limitations}

While our in-context learning strategy effectively constrains the LLM to the correct API syntax, the inherent stochasticity of foundation models means that code generation is not deterministic. In rare instances, the model may hallucinate non-existent functions, necessitating a regeneration step, especially when requests drastically diverge from provided documentation. Additionally, like many particle-based solvers, the underlying physics engine can exhibit numerical instability when subject to extreme physical parameters (e.g., excessively high stiffness or velocity). Finally, our current use of convex hull proxies, while highly efficient for real-time interaction, limits the simulation fidelity for objects with deep concavities or complex topology.
\section{Conclusion}

We introduced \ours, a text-to-physics framework that resolves the longstanding divide between open-vocabulary language control and physically grounded 3D Gaussian animation. By translating natural language prompts into executable code for an efficient physics engine, our approach sidesteps the limitations of text-driven Gaussian manipulation methods, namely the lack of physical grounding and the rigidity of hand-authored simulation setups. Experiments across diverse scenes and prompts show that \ours achieves higher perceptual alignment, better motion quality, and greater realism than competitors, while operating at runtimes that support interactivity and iterative editing. This paradigm opens a promising path toward general, physically plausible, and authorable 3DGS animation tools. We believe that this paves the way to future works exploring richer scene-level interactions, more complex multi-object dynamics, and broader extensions of physics-based 3DGS animation.
{
    \small
    \bibliographystyle{ieeenat_fullname}
    \bibliography{main}
}
\clearpage
\setcounter{page}{1}
\maketitlesupplementary

\section{Video Supplementary Material}
We encourage readers to watch the accompanying video results, as we focus on the temporal animation of Gaussians, which is best appreciated in motion. In particular:

\noindent(1) we include a comparison between baselines and PhysTalk on all prompts evaluated in Fig.~\ref{fig:qualitative_comparison}. This allows one to appreciate the elastic dynamics produced by PhysTalk and its behavior when generating dense lava flows;

\noindent(2) we show the additional qualitative results provided in Fig.~\ref{fig:additional_qualitative}. These highlight the reduced moon gravity and the behavior of multimaterial objects, where the rigid part remains undeformed while transferring impact forces to the elastic part, which then wobbles;

\noindent(3) we assess the fluid dynamic behavior with and without the hole filling strategy described in Sec.~\ref{sec:phy_and_ski}: when hole filling is active, the lava appears denser and forms a continuous stream, rather than showing visible gaps;

\noindent(4)  we illustrate both the simulation capabilities of our pipeline and its editability by reusing simulations from Fig.~\ref{fig:qualitative_comparison} and varying their physical parameters: we tune both the Young's modulus $E$, which controls objects' elasticity, and the parameter $\gamma$, that is the surface tension, which controls the spread of the lava;

\noindent(5) we show a user interacting with both the vase and the bulldozer scenes. Here yellow arrows point in the direction of the force applied by the user. Also, forces are proportional to the coordinate system's displacement. Notably, objects react to pushes and gravity, while respecting the elastic material properties chosen in the simulation.

\section{Simulation Grounding}

In this Supplementary Material we provide the complete simulation grounding documents.

\noindent\textbf{Few-Shot Exemplars.} We constrain the LLM to generate executable Genesis code using a small set of hand-written few-shot exemplars, described below.

\noindent\texttt{rigid.py}: this simulation describes how to simulate a rigid body falling. This is not trivial, as unlike elastic, liquid, muscle materials, rigid bodies in Genesis do not leverage a particle system. This prevents us from directly obtaining the data required to skin the simulated motion back to the Gaussians described in Sec.~\ref{sec:phy_and_ski}. Hence, the rigid object's motion is transferred to a set of evenly distributed particles that we spawn inside the simulated convex hull. 

\noindent\texttt{turnToWater.py}: this describes how to turn an object into a very sparse fluid that quickly spreads over a flat surface.

\noindent\texttt{multimaterial.py}: this simulation describes how to construct an object with an elastic bottom part and a rigid top part. This is crucial as Genesis does not natively support these objects. We contribute a procedure to create them, and it is essential to instruct the LLM on how to use it. 

\noindent\texttt{drop.py}: this simulation describes how to drop an object and make it fracture at impact into fragments that are approximated as sand-like particles.

\noindent\textbf{APIs}. In \texttt{common.py} we provide the LLM with a set of functions for recurring operations, including:

\noindent\texttt{getConvexHull}: this function provides an easy and efficient interface to create watertight convex hull meshes from Gaussian centers. 

\noindent\texttt{get\_rot}: this function allows the LLM to pass a Genesis entity and obtain  its deformation gradient $F$. In the case of fluids, for which the deformation gradient is not defined, it returns the identity matrix $I$. 

\noindent\texttt{deform\_centers\_and\_rot}: this function computes a mapping between simulation particles at rest position and the original object's Gaussian centers. The mapping is computed at the first invocation and then cached. For each Gaussian it selects $K$ neighboring particles and their deformation gradients, then applies inverse-distance blending to obtain the Gaussian's motion and deformation.

\noindent\textbf{Instructions.} In Fig.~\ref{inst_1}–\ref{inst_3} we report the instruction used to steer the LLM toward generating valid Genesis simulation code. These instructions have been tuned to be sufficiently general while fostering reproducibility of the qualitative results provided in the main paper. 

\section{Additional Details}

Finally, we report some additional details about our pipeline. While in our experiments we use GPT-5 as the LLM and Genesis as the physics engine, we emphasize that the overall design is modular and these components could be replaced with any retrieval augmented LLM and any particle based physics engine. For instance, although we currently manually fit a box and a ground plane under the horse to sustain the lava flow, one could incorporate a scene segmentation module that automatically provides a semantic and geometric description of the scene. 
We also instruct the model to default to an elastic material when not otherwise specified, since elastic motion is easier to appreciate than rigid motion in sequences of static images. Also, as the LLM is a generative model, it may produce different simulations for the same prompt.
Finally, before creating the convex hull, we prune Gaussians that have few neighbors within a local spherical neighborhood. This step is crucial, since outliers due to imperfect reconstruction can lead to distorted convex hulls and degraded physics simulations.

\clearpage

\begin{figure*}[p]
  \centering
  \vspace{-0.5em}
\begin{lstlisting}[firstline=1,lastline=50]
Generate simulate functions or a Simulation class for Genesis-AI that animate elements using physics. Always rely on the provided documentation, user guide, API guide, and example files. Do not invent APIs.

Rules
1) Use official APIs only. Before calling any function, verify it exists and confirm its signature, parameter names, types, and return values in the provided documentation and examples. Always read the documentation.
2) No hallucinations. If an operation is not supported by Genesis or not present in the examples, do not use it.
3) Physically plausible setups. Default to real-world behavior, including gravity and contact, unless the prompt clearly implies otherwise.
4) Never re-define, alter, or re-implement anything from common.py, especially shutdownGS, deform_object, deform_centers_and_rot, get_rot, getConvexHull.
6) Provide code in a code-box so that it can be easily copied and pasted
7) Be sure that the physical parameters (rho, dt, viscosity etc...) are not too high, otherwise the simulation will explode.
8) Do not use "very thin clearance to avoid initial interpenetration"
9) No cameras

Required File Header (exact)
At the very top of every generated simulation function or Simulation class, emit this header verbatim, replacing [PROMPT] with the user's prompt:
#Prompt: [PROMPT]
import genesis as gs
import numpy as np
from math import floor
import torch
from .common import shutdownGS, deform_centers_and_rot, get_rot, getConvexHull

Class Signature & Public API
- Never change this constructor signature:
class Simulation():
    def __init__(self, pts=None, fps=50, s=5, use_birdal_quat=True, use_convex_hull=True):
- Always include these public methods with the same names and roles:
  step(n_substeps=1): advance the simulation, sample exactly one frame per step() call.
  query(clear=True): return deformed centers and F_deform up to now, then clear buffers when clear is True.
  shutdown(): call shutdownGS(self.scene).
- Sampling buffers:
  Maintain self.positions and self.F_deform lists.
  When exporting accumulated data inside query, you must concatenate with these exact lines, do not change them:
    simulations = np.concatenate(positions, axis=0)
    F_deform = np.concatenate(F_deform,axis=0)
- always use a self._build_scene() where you do gs.init(backend=gs.gpu) and the set up the scene
- notice that in gs.morphs.Mesh(file=value ...) the value must be string, not posix path

Data Collection Details
- Do not write positions.append(mpm_sand_box.get_particles()).
- Use this pattern to read particle positions from MPM or SPH entities:
  positions.append(mpm_elm.solver.particles.to_numpy()['pos'].copy()[0].squeeze(1))
  or the equivalent shape-correct version already used in the examples.

Ground Plane and Initial Placement
- Every simulation must spawn a ground plane under the object at its original pose, and the object's bounding box must be aligned to this pose.
- If asked to start the box from a different pose, adjust after the ground plane is created and before the box is instantiated. For example, to raise by 1 m:
  pos[2] += 1
  pts[:, 2] += 1
  cube = scene.add_entity(
      morph = gs.morphs.Box(pos=pos, size=size),
\end{lstlisting}
  \vspace{-1em}
  \caption{Instructions (lines 1-50)}
  \vspace{+1em}
  \label{inst_1}
\end{figure*}

\begin{figure*}[p]
  \centering
  \vspace{-0.5em}
\begin{lstlisting}[firstline=51,lastline=100]
  )
The plane is always:

self.scene.add_entity(
            gs.morphs.Plane(pos=(pos[0], pos[1], plane_z)),
            gs.materials.Rigid(friction=.15),
            gs.surfaces.Rough(color=(.25,.25,.25)))

If asked to float or apply an initlial force or push in direction x to the object you can use elastic entity and do:
        self.scene.build()
        vel = torch.zeros((self.cube.n_particles, 3), dtype=torch.float32)
        vel[:, x] = some_value
        self.cube.set_velocity(vel)
Velocity values around 5 are very strong forces. 4 is good.
Do not use undocumented DOF-velocity controls for a free rigid.

Materials, Stability, and Samplers
- Choose materials reflecting prompt and remain numerically stable. Avoid extreme densities, stiffness, or particle sizes that cause explosions or collapse.
- For SPH liquids valid samplers are: regular, pbs, pbs-sdf_res. Do not use poisson for gs.materials.SPH.Liquid.
- Use conservative dt, substeps, particle size, and friction to avoid instability. Prefer parameters demonstrated in the example files.
- The maximum mu is 0.17, don't use higher values.  
- Very dense fluids are controlled by gamma and have gamma = 0.4. 
- If asked for elastic materials use gs.materials.MPM.Elastic with E close to 1.0e5. model="neohookean" is not given for elastic models, and also damping keyword do not use it. Only in case of multimaterial you can use Muscle and give it self.E_soft = 6. For the rigid part use rho_rigid=500.0 .
- 'MPMEntity' object has no attribute 'set_friction'
- If asked for rigid object use rigid_box_fakecloud,rigid_pose_as_frames provided rigid.py
- If rigid or liquid isn't explicitly specified in prompt by user use mpm.elastic for objects

Prompt Interpretation
- If prompt is like turn x into y, make x into y, or make x do this, create a simulation that makes the object perform the requested action using physically meaningful materials and forces.
- Map high-level text to concrete physics choices. Example, turn into lava implies SPH liquid with reasonable density and viscosity. Drop from 1 m implies raising initial pos and pts by 1 m before creating the box entity.

Convex Hulls and Geometry
- Represent the box-like object either as:
  a) A convex hull built from pts using getConvexHull(pts) when use_convex_hull is True, or
  b) A parametric gs.morphs.Box(pos=pos, size=size) when use_convex_hull is False.
- Always compute aabb_min, aabb_max, pos, and size from pts in the constructor or a builder routine.

Skinning and Output
- During stepping, record particle positions and F_deform and append as described above.
  You can obtain Position and F_deform via get_rot(entity) and append as numpy arrays. If a tensor, detach to CPU first.
- In query, call use deform_centers_and_rot like:
  centers, F,  mapping = deform_centers_and_rot(
              simulations, F_deform, self.pts, mapping=self.mapping, use_birdal_quat=self.use_birdal_quat)
        
  Maintain and reuse self.mapping once it is populated to keep k-NN consistent.

Query always returns centers, F,  self.liquid_resample

Multimaterial Hybrid Objects
- For multimaterial object follow the multimaterial.py, where upper 50% is rigid, lower is muscle-elastic. Change the materials of these parts if requested by user. 
\end{lstlisting}
  \caption{Instructions (lines 51-100)}
\end{figure*}

\begin{figure*}[p]
  \centering
  \vspace{-0.5em}
\begin{lstlisting}[firstline=100,lastline=151]
- For multimaterial object follow the multimaterial.py, where upper 50% is rigid, lower is muscle-elastic. Change the materials of these parts if requested by user. 
-For multimaterial elastic means muscle with nu around 0.005, model="neohooken", n_groups=2
-use rho_soft around 1,  rho_rigid around 500.0 .
- mu_ground= 0.35  mu_block= 0.35 are allowed for multimaterial
- When the prompt says "lower part rigid, upper part elastic," always use the lower segment's geometry to create the rigid URDF (with proper local frame and no unintended offsets), use the upper segment's geometry for the soft body, and apply any small vertical gap to the soft upper piece above the rigid base to prevent overlap.
- use tilt_euler_deg=(8,5,0) and uses R_tilt to offset the soft center by gap along local +z.
- computes (trans0, quat0) via gu.transform_pos_quat_by_trans_quat(geom.init_pos, geom.init_quat, base.init_x_pos, base.init_x_quat) - i.e., transforms the geom's local pose by the link's initial pose

Safety and Stability Checklist, apply before finalizing
- Time step dt, substeps, and particle size are moderate and similar to examples.
- Material parameters are within realistic ranges for the described effect.
- The first step() call samples at most one frame into buffers.
- positions and F_deform have consistent shapes for concatenation.
- The ground plane exists and sits under the initial object AABB.
- SPH sampler, if used, is regular, pbs, or pbs-sdf_res.
- No function from common.py has been modified or re-implemented.
- The class signature, method names, concatenation lines, and import header are exactly as specified.
\end{lstlisting}
  \caption{Instructions (lines 100-116)}
  \label{inst_3}
\end{figure*}

\end{document}